\documentstyle[11pt]{article}
\newcommand{\dl}{\delta^{(3)}({\bf r})}

\newcommand{\BE}{\begin{equation}}
\newcommand{\EE}{\end{equation}}
\newcommand{\BA}{\begin{eqnarray}}
\newcommand{\EA}{\end{eqnarray}}

\begin{document}
\begin{titlepage}

\vspace*{1mm}
\begin{center}

            {\LARGE{\bf A connection between gravity  \\ 
                        and the Higgs field }}

\vspace*{14mm}
{\Large  M. Consoli }
\vspace*{4mm}\\
{\large
Istituto Nazionale di Fisica Nucleare, Sezione di Catania \\
Corso Italia 57, 95129 Catania, Italy}
\end{center}
\begin{center}
{\bf Abstract}
\end{center}

Several arguments suggest that 
an effective curved space-time structure (of the type
as in General Relativity) can actually find
its dynamical origin in an underlying condensed medium of spinless
quanta. For this reason, we exploit
the recent idea of density fluctuations in a `Higgs condensate' with the 
conclusion that such long-wavelength effects might 
represent the natural dynamical agent of gravity. 
\vskip 35 pt
\end{titlepage}

\section{Introduction}

Following the original induced-gravity models \cite{adler}, one may
attempt to describe gravity as
an effective force induced by the vacuum structure, in analogy
with the attractive interaction among electrons that can only
exist in the presence of an ion lattice. 
If such an underlying `dynamical' mechanism is found, 
one will also gain a better understanding of
those peculiar `geometrical' properties that are the object 
of classical General Relativity. In fact, 
by demanding the space-time structure to 
an undefined energy-momentum tensor, no clear distinction 
between the two aspects is possible and this may be the reason for 
long-standing problems (e.g. the space-time curvature associated
with the vacuum energy).

A nice example to understand what aspects may be
kinematical and what other aspects may be dynamical 
has been given by Visser
\cite{visser}. Namely, a curved space-time pseudo-Riemannian geometry
 arises when studying the density fluctuations in a moving
irrotational fluid, i.e. where the velocity field is the gradient of a
scalar potential $\sigma (x)$. In this system, the underlying space-time is 
exactly flat but the propagation of long-wavelength
fluctuations is governed by a curved 
 `acoustic' metric 
$g_{\mu\nu}(x)$ determined at each space-time point $x$ 
by the physical parameters of the fluid
(density, velocity and pressure). These can all be expressed in terms
of $\sigma(x)$ through the hydrodynamical equations.
In this example, there is a system, the fluid, whose 
constituents are governed by some underlying molecular
interactions that, on macroscopic scales, 
can be summarized in the value of
a scalar field $\sigma(x)$. At some intermediate level, 
this contains  the dynamical informations. On the other hand, 
the effective metric tensor
$g_{\mu\nu}$ is a purely kinematical quantity that depends on $\sigma(x)$ 
in a general parametric form
\BE
\label{parametric}
         g_{\mu \nu}(x)=g_{\mu \nu}[\sigma(x)]
\EE
 Notice that there is no
non-trivial curvature in the equilibrium state $\sigma(x)=\sigma_o={\rm const.}$
where any fluid is {\it self-sustaining} \cite{volovik}. 

Looking for the physical origin of gravity, one should also look
at ref. \cite{barcelo}.
It contains an exhaustive set of many gravity-analogs, 
namely of systems that can 
simulate and/or reproduce the experimental properties of gravitation (moving
fluids, condensed matter systems with a refractive index,  
Bose-Einstein condensates,..). Again
the hydrodynamical analogy, when exploited in a lagrangian
field-theoretical context, leads to 
the emergence of a curved, effective 
geometry as a result of linearizing a scalar
field theory (in flat space-time)
around some non-trivial background configuration. For this reason, one 
may agree with the authors of ref.\cite{barcelo} and 
conclude that General Relativity provides
some kind of universal
low-energy picture, just as hydrodynamics. This, concentrating on the
properties of matter at scales that are 
much larger than the mean free path for the elementary
constituents, is
insensitive to the details of short-distance molecular dynamics. 
This point of view should not sound too surprising to the extent that is
consistent with the historical origin of General Relativity before the birth
of quantum mechanics. 

Notice that we are not saying that one can {\it prove} a formal 
equivalence with Einstein's General Relativity.
In fact, a medium 
provides a definite physical framework whose
short-distance details might have no counterpart
in General Relativity. However, by restricting to 
long-wavelength phenomena, and
to some degree of accuracy, the two descriptions may be indistinguishable.
At the same time, there may be different
dynamical scenarios at very small scales that, however,
become equivalent to General Relativity on a larger scale.
Their number can only be restricted by exploiting this type of
correspondence, within the constraints of the experimental results. 
For instance, $\sigma(x)$, rather than arising from a fundamental 
scalar theory, might turn out to be 
a collective excitation of a superfluid fermionic vacuum
\cite{volovik}.

On the other hand, the idea of an `Aethereal Medium', whose density variations
could account for the gravitational force, is deeply rooted in the origin of
our scientific culture and can be found, for instance, in Newton himself. 
In fact, although refraining from a definite hypothesis on
the physical origin of gravity in the Principia 
(".. I frame no hypotheses..and to us it is enough that gravity does really 
exist, and act according to the laws which we have explained .."), 
Newton explicitely inserted some additional 
{\it Queries on the Aether} in the last version of Opticks.
 In particular, in Query
21, he was considering that "...if the elastick force of this Medium be
exceeding great, it may suffice to impel Bodies from the denser parts of the
Medium towards the rarer, with all that power which we call Gravity". To this
end, the hypothetical aetherial particles should feel a repulsive force 
("...Particles which endeavour to recede from one another..") and be
"...exceedingly smaller than those of Air, or even than those of Light..."
thus making "...that Medium exceedingly more rare and elastick than Air..".

It should not be too difficult, at least
for a present-day particle physicist, to 
realize that the required properties of this hypothetical medium might have
a well defined counterpart in the quantum vacuum of 
a spontaneously broken $\lambda\Phi^4$ theory: 
the `Higgs condensate'.  In fact, the name itself means that 
a non-vanishing expectation value 
$\langle \Phi \rangle \neq 0$ should
correspond to some kind of medium, an aether, 
made up by the physical condensation process of elementary spinless quanta, 
the `phions' \cite{mech}, whose `empty' vacuum state is not the true ground
state of the theory. 

The symmetric phase, where $\langle \Phi \rangle=0$, 
will eventually be re-established at a phase
transition temperature $T=T_c$. This, in the Standard Model, is so high 
that we can safely approximate the scalar condensate as a zero-temperature 
system. This observation provides the argument to
represent the phion condensate as a medium
where bodies flow without any apparent friction, as superfluid $^4$He.
In fact, a zero-temperature Bose system is a `quantum liquid', i.e. a system
whose macroscopic properties depend
on the quantum nature of its constituents. This requires
a form of quantum hydrodynamics, of the type originally considered
by Landau \cite{hydro}, where the local density of the fluid 
$n({\bf{r}})$ and 
the current density vector ${\bf{J}}({\bf{r}})$ 
have canonical commutation relations \cite{pita}, 
 as in quantum mechanics for the position
and momentum operators. 

The analogy with superfluid helium is also supported by the other
observation \cite{mech} that, as for the 
interatomic $^4$He-$^4$He potential, the low-energy limit of cutoff
$\lambda\Phi^4$ is 
a theory of spinless quanta with a short-range repulsive core and 
a long-range attractive tail. 
The latter originates from ultraviolet-finite parts
of higher loop graphs \cite{mech} that give rise to a 
$- {{\lambda^2 e^{-2m_\Phi r} }\over{r^3}}$ attraction where $m_\Phi$ is the
phion mass. Differently from the usual ultraviolet divergences, 
this ultraviolet-finite part cannot be reabsorbed into a standard 
re-definition of the tree-level, repulsive $+\lambda\dl$ 
potential and is essential for a physical description of the
condensation process 
when approaching the phase transition limit $m_\Phi \to 0$ where the symmetric 
vacuum at $\langle\Phi\rangle=0$ becomes unstable. 

Although representing some kind of aether, the scalar condensate is,
however, different from the aether of classical physics and
there are phenomena that cannot be
understood at the classical level. For instance, since the superfluid flow is 
a potential flow, there is no drag force on a body moving in the fluid and 
no D'Alembert paradox \cite{fluid} as if
the fluid were locally dragged by all moving bodies.
In this sense, 
an origin of gravity from the phenomenon of vacuum condensation 
would require only a modest switch from 
Einstein's Special Relativity to Lorentzian Relativity, with 
the local gravity field providing the operative definition of Lorentz's aether
\cite{tomgps}. 

This point of view is consistent 
with the {\it correct} interpretation of stellar 
aberration. To this end, one should look at refs.\cite{ives,phipps}.
In particular, in Phipps's paper \cite{phipps}, that starts from the original 
De Sitter's remark that double star systems with high relative velocities show
the same aberration angle, it is clearly pointed out that the experimental value
$\theta\sim 20.5"$ with $\tan\theta \sim {{v}\over{c}}$, defines a 
velocity $v \sim 30$ km/sec that is {\it not} the 
relative velocity between the observer and the emitting source. 
Rather, $v$ is associated with the 
motion of the observer, in our case the earth, in the gravitational 
field of the sun. The same conclusion is suggested by Synge \cite{syngephipps} 
who states that for observational purposes, the two frames S and S' (connected
by the Lorentz transformation with velocity $v$) "...consist of the earth 
himself at two different positions in its orbit around the sun". This implies
that an hypothetical observer placed on Pluto, would
observe an aberration angle $\theta\sim 3.3"$ and that another hypothetical 
observer placed on the Sun would observe no aberration at all. This last
conclusion is indeed consistent with Bergman's treatment \cite{bergman} where
the aberration angle relates "...the direction of the incoming light with
respect to two frames of reference, that of the sun and that of the earth".

Therefore, since De Sitter's remark dates back to 1913, one can understand 
why the idea of General Relativity as a description of the 
deformations of a peculiar, aethereal medium was seriously considered by 
Einstein in the period 1916-1924. This `resurrection of aether' in Einstein's
mind is confirmed by a considerable amount of published and
unpublished manuscripts reported by Kostro
\cite{kostro}, including the famous Leyden lecture. 
According to this original picture, the inclusion of gravity, and therefore the
transition from Special to General Relativity, could be understood by 
replacing the aether of classical physics with a `sublimated' 
aether whose constituents do not follow definite space-time trajectories: 
today we would say `that have a quantum behaviour'. 
However, independently of
quantum mechanics, looking at ref.\cite{kostro}, one will
discover that there might have been several reasons, even quite 
unrelated to physics, why the `sublimated' aether
was not exploited by Einstein in more detail. 

At the same time, just 
the quantum phenomena might induce to a change of perspective
in the approach to `geometrize' gravity.
This, historically, started from experimental 
observations, namely the universality of free fall. However, other 
experiments show that quantum mechanical wave functions are very
different from the geometrical objects of classical tensor calculus
\cite{hollandg}. In fact, one can explicitely prove \cite{greenberger} 
the `strong' equivalence 
between a gravitational field and an accelerated frame. The price to pay
is to admit mass-dependent phase transformations of the wave functions.
Therefore, the possibility of re-absorbing the effects of gravity 
into a universal change of space-time might simply reflect
the net cancellation of the quantum interference effects
in the classical limit. This might be another indication for
General Relativity being an effective theory which averages over
distances that are much larger than the 
typical atomic sizes. 

In this spirit, we shall start in Sect.2 
from the basic properties of curved space-time where,  
for long wavelenghts, the excitations of our medium can be described 
in terms of a single scalar function
$\sigma(x)$. This is {\it experimentally} known to be
the Newton potential. Its possible 
interpretation in terms of collective density fluctuations of the 
Higgs condensate will be proposed 
in Sect.3. Finally, Sect.4 will contain our summary and a discussion of
some general consequences of our approach.

\section{Curved space-time and a `medium' with a scalar field}

Describing the
possible excitation states of a superfluid medium
is an extremely difficult task. 
In this section, we shall adopt a minimal point of view, namely trying to
 re-absorb its properties into a single
 scalar function
\BE
\Phi(x) \equiv e^{-\sigma(x)} \langle \Phi \rangle
\EE
that deviates from its equilibrium constant value corresponding to $\sigma=0$. 
 As anticipated in the Introduction, this type
of description should work in the
hydrodynamical regime, when the wavelengths associated with the
fluctuations are much larger than the mean free path for the
elementary constituents. 

As a possible form of interaction between the fluctuations of the medium
and ordinary matter, we shall adopt the simple Higgs model, namely a coupling
to the particle mass with the replacement
\BE
 m_o \to     m_o{{\Phi(x)}\over{\langle\Phi\rangle}}=
m_o e^{-\sigma(x)}
\EE
As anticipated, we shall also assume that $\sigma(x)$ is a very slowly
varying function such that one can completely neglect its variation over distances
that are comparable with the atomic size. In this situation, 
the effect of a non-zero $\sigma$ on the energy levels of 
a hydrogen-like atom simply amounts to a 
re-definition of the electron mass with an average constant value
$m= m_o e^{-\sigma}$ in  the Dirac Hamiltonian 
\BE
\label{dirac}
        H_D= {\bf{\alpha}}\cdot{\bf{p}} + \beta {m} -{{Ze^2}\over{r}}
\EE
This changes the energy levels and the
frequencies $\omega_o \to \omega_o e^{-\sigma}$. 
 Therefore, the natural 
period of an atomic clock $T={{2\pi}\over{\omega}}$ is changed, 
$T=T_oe^{\sigma}$, with 
respect to the value $T_o={{2\pi}\over{\omega_o}}$ associated with $\sigma=0$.
Analogously, the Bohr radius $r_B={{\hbar}\over{ Ze^2 m_o }}$ is changed into
 $ r_B e^{\sigma} $ thus producing a symmetric re-scaling of
 the length of the rods. Since all masses are affected in the same way, and
the units of length and time scale as inverse masses, the overall effect 
is equivalent to a conformal 
re-scaling of the metric tensor
\BE
\label{conformal}
                 g_{\mu\nu}(x)=e^{2\sigma(x)}\eta_{\mu\nu}
\EE
where $\eta_{\mu\nu}$ denotes the Minkowski metric.

However, the idea of 
$\sigma(x)$ as the fluctuation of a (`non-dispersive')
medium, suggests
another physical effect: the introduction of a refractive index. In this case, 
before the conformal re-scaling, the Minkowski metric would be replaced by
\BE
\label{index0}
\hat{\eta}_{\mu\nu}\equiv({{1}\over{ {\cal N}^2}},-1,-1,-1)
\EE
with a refractive index ${\cal N}={\cal N}(\sigma)$ and a normalization
 such that ${\cal N}=1$ when $\sigma=0$
(when no confusion can arise, we shall set to unity
the Newton constant $G_N$ and the speed of light in the `vacuum' $c$).

 Thus we obtain the metric structure
\BE
\label{basic1}
         \hat{g}_{\mu \nu}\equiv 
({{e^{2\sigma}}\over{{\cal N}^2}},-e^{2\sigma},-e^{2\sigma},-e^{2\sigma})
\EE
that re-absorbs the local, isotropic modifications of Minkowski 
space into its basic ingredients: the value of the speed of light and 
the space-time units. Equivalently, the same metric structure can be 
interpreted as arising from {\it separate}
local changes of the space and time units. 
In fact, such a transformation is known
to represent one of the many possible ways, perhaps the most fundamental, 
to introduce the concept of curvature \cite{feybook,dicke1, szondy}.
In particular, there exist definitions of units, depending
on a scalar field, for which 
a general curved space-time becomes flat, all the Riemannian invariants 
being zero \cite{dicke2}. 

The idea of introducing a refractive index
is very natural (see e.g. \cite{rosen}) at least when comparing with 
any known medium with definite physical properties. For instance, 
when considering, as a possible model,
the non-trivial vacuum of an underlying scalar quantum field theory, we get
the simple picture of a condensate of spinless quanta. If these are treated as 
`hard-spheres', the physical properties of the vacuum would depend
on their scattering length and particle density. On this 
basis, Lenz \cite{lenz} first showed that such a system behaves like a medium
with a refractive index. 

Without attempting a microscopic derivation, we can try to deduce
a possible form of ${\cal N}(\sigma)$ using some general arguments. 
To this end, we first observe that, for a time-independent
 $\sigma(x)$, our metric Eq.(\ref{basic1}) 
is just a different way to write the general isotropic metric
\BE
\label{isotropic}
 \hat{g}_{\mu \nu} \equiv (A,-B,-B,-B)
\EE
Now, one may ask when the 
local light velocity $c(x,y,z)\equiv \sqrt{ {{ A}\over{B}} }$, 
 defined as a `particle' velocity 
from the condition $ds^2=\hat{g}_{\mu\nu} dx^\mu dx^\nu=0$, agrees with the 
curved-space equivalent of the
phase velocity of light pulses. These are
solutions of  the D'Alembert wave equation with the metric $(A,-B,-B,-B)$
\cite{progress}
\BE
\label{dalembert}
{{1}\over{A}}
{{\partial^2 F }\over{\partial t^2}}
-{{1}\over{B}}(
{{\partial^2}\over{\partial x^2}}+
{{\partial^2}\over{\partial y^2}}+
{{\partial^2}\over{\partial z^2}})F -
{{1}\over{\sqrt {AB^3} }} (\nabla \sqrt{AB})  
\cdot (\nabla F)=0 
\EE
so that, by introducing the 3-vector ${\bf{g}}\equiv 
\sqrt { {{A}\over {B^3}} } (\nabla \sqrt{AB})$ we obtain 
\BE
\label{omega}
{{\partial^2 F }\over{\partial t^2}}
              = {{A}\over{B}}~ \Delta F + {\bf{g}}\cdot (\nabla F)
\EE
By identifying 
${{1}\over{F}}{{\partial^2 F }\over{\partial t^2}}$ as the equivalent 
of $-\omega^2$ and ${{1}\over{F}}\Delta F$ as the corresponding of $-k^2$, 
we find that
particle velocity and phase velocity
$c_{\rm ph}\equiv {{\omega}\over{k}}$ 
agree with each other only when ${\bf{g}}=0$, i.e. when
$AB$ is a constant. This product 
can be fixed to unity with flat-space boundary 
conditions at infinity and, therefore, the resulting value 
\BE
\label{index}
{\cal N}=e^{2\sigma}
\EE
can be considered a consistency requirement on our medium 
to preserve
the observed particle-wave duality which is intrinsic in the nature of light. 
On the other hand, if ${\cal N} \neq e^{2\sigma}$, we should specify the operative 
definition used for the local speed of light: a) 
the time difference for a light pulse to go forth and back between
two infinitesimally close objects at relative rest, b)
the value obtained combining frequency and wavelength of a given radiation
source,..

With this choice, Eq.(\ref{basic1}) becomes
\BE
\label{basic2}
         \hat{g}_{\mu \nu}\equiv
(e^{-2\sigma},-e^{2\sigma},-e^{2\sigma},-e^{2\sigma})
\EE
and could be used 
for a space-time description in a medium specified by a
given scalar field $\sigma(x)$.

Alternatively, if we assume that our $\sigma(x) \neq 0$ is equivalent 
to a gravitational field, we can compare the metric structure Eq.(\ref{basic2}) 
with the experiments in the known 
weak gravitational fields where the gravitational potential does not change
appreciably over distances, say, of a few millimeters. 
In this case, we obtain the experimental result
that $\sigma$ is just the Newton potential, i.e.
(for a centrally symmetric field) 
\BE 
\label{sigma11}
               \sigma_{\rm exp}= {{M}\over{r}}
\EE
As anticipated, in this description, the space-time curvature is equivalent to
suitable local re-definitions of the space and time units. Thus, for instance, 
the gravitational red-shift is explained through the behaviour of clocks 
in the gravitational field whereas the energy of the propagating photon 
does {\it not} change with the height \cite{okun} and there is no need to 
introduce the concept of an effective gravitational mass for the propagating 
photons. Analogously, the deflection of light 
can be explained by converting the units of time
and length that define the local speed of light into those that are used 
for the plotted speed \cite{schiff}.  

Let us now compare our `medium' with 
General Relativity. To this end, we shall study the class of
metrics that are solutions of the following field equations
\BE
\label{general}
 R_{\mu\nu}-{{1}\over{2}}g_{\mu\nu}R
= \lambda (\sigma_\mu \sigma_\nu -{{1}\over{2}} g_{\mu\nu} 
\sigma^\alpha\sigma_\alpha)
\EE
These can be considered Einstein equations in the presence of 
an energy-momentum tensor for a scalar field $\sigma(x)$, 
for various values of the parameter $\lambda$. For a centrally
symmetric field $\sigma=\sigma(r)$, 
Tupper \cite{tupper} has shown that {\it all} solutions of Eqs.(\ref{general})
 are 
consistent with the three classical weak-field tests and the
   Shapiro planetary radar reflection experiment. 
Depending on the value of the parameter 
$\lambda$ one gets the Schwarzschild metric, 
for $\lambda=0$, or the Yilmaz metric \cite{yilmaz1}, 
for $\lambda=2$. For any $\lambda$, the solutions 
 are conformal transformations of solutions
of the Brans-Dicke theory \cite{brans}.

 However, only for $\lambda=2$ the resulting
metric tensor depends {\it parametrically} on $\sigma$. In this case, 
Eqs.(\ref{general}) become algebraic {\it identities} 
consistently with the point of
view Eq.(\ref{parametric}) that the
metric tensor is an auxiliary kinematical quantity 
depending parametrically on more fundamental fields. 
In this case, for $\lambda=2$, where
$\sigma= {{M}\over{r}}$ represents the Newton potential, 
the Yilmaz metric in isotropic form (`Y'=Yilmaz)
\BE
\label{yilma2}
g^{Y}_{\mu\nu}\equiv (e^{-{{2M}\over{r}}} , 
-e^{ {{2M}\over{r}}}, -e^{{{2M}\over{r}}}, -e^{{{2M}\over{r}}})
\EE
is formally identical to Eq.(\ref{basic2}).

Notice that the idea of the Newton potential as a fundamental 
scalar field is {\it not} restricted to flat space but 
has a meaning also in curved space-time. Clearly, 
adopting the point of view of a parametric dependence of the metric tensor
is equivalent to say that the `right' choice among all possible values
of $\lambda$, is just $\lambda=2$, i.e. the old Yilmaz theory.
Finally, having an independent argument 
for $\sigma(x)$ being the Newton potential, we 
could describe the relevant space-time properties without solving
any field equation. 

The choice $\lambda=2$ is also suggested
by the Yilmaz remark \cite{remark} that the metric Eq.(\ref{basic2}), 
 differently from all other 
metrics, is a solution of Einstein field equations, 
even for a gravitational many-body system where 
 $T_{\mu\nu}\equiv(\rho,0,0,0)$ and
\BE
\label{rho}
         \rho({\bf{x}} ) = -4\pi \sum_n M_n
\delta^3( {\bf{x}} - {\bf{x}}_n) 
\EE
Indeed, when replacing the Newton potential 
\BE
\label{remark}
         \sigma({\bf{x}} ) =  \sum_n {{M_n}\over
{ |{\bf{x}} - {\bf{x}}_n|}} 
\EE
in the metric Eq.(\ref{basic2}), the field equations 
\BE
\label{general1}
 R_{\mu\nu}-{{1}\over{2}}g_{\mu\nu}R
= 2 (\sigma_\mu \sigma_\nu -{{1}\over{2}} g_{\mu\nu} \sigma^\alpha\sigma_\alpha)
+2 T_{\mu\nu}
\EE
become
algebraic identities. Therefore, if one 
wants to compare with real many-body gravitational systems, 
Eq.(\ref{basic2}) represents a very convenient starting point for any
time-dependent approximation. Indeed, for slow motions, the situation
is similar to the conventional adiabatic Born-Oppenheimer 
approximation where one approaches the time-dependent problem
by expanding in the eigenfunctions of a
static 2-center, 3-center,.. n-center hamiltonian.
Actually, by inspection of the second-order terms, 
Yilmaz claims \cite{claim} that
the standard 
Einstein-Hoffman-Infeld metric, used for a description
of classical motions in the solar system, represent 
a time-dependent improvement on the static solution of 
Eq.(\ref{general1}) with $\sigma(x)$ as in Eq.(\ref{remark}).
Namely, the scalar field would be implicitely taken into account.
Without committing ourselves to this 
particular aspect, we shall return, however, to a comparison 
with the Schwarzschild metric in the conclusions. 

Before concluding, we observe that the metric structure Eq.(\ref{basic2})
was also obtained by Dicke \cite{varenna} in a stimulating remake of 
Lorentz's electromagnetic 
aether. This is based on the simultaneous replacements of the particle mass
\BE
\label{replace1}
                      m_o \to m_o f (\epsilon,\mu)
\EE
and of the light velocity
\BE
\label{replace2}
                      c^2 \to {{c^2}\over{ \epsilon \mu}}
\EE
where $\epsilon$ and $\mu$ are respectively the space-time dependent
dielectric function and 
magnetic permeability of the aether. Consistency with 
the experimental results (the E\"otv\"os experiment, velocity independence of
the electric charge,...) requires $\epsilon =\mu$ leading to the effective
metric structure (`LD'=Lorentz-Dicke)
\BE
\label{lorentz}
 g^{\rm LD}_{\mu\nu}\equiv ({{f^2}\over{\epsilon^4}},
-{{f^2}\over{\epsilon^2}},
-{{f^2}\over{\epsilon^2}},
-{{f^2}\over{\epsilon^2}})
\EE
Finally, a comparison with the classical tests in a weak gravitational field
gives
\BE
\label{ff}
                 f^2= \epsilon ^3
\EE
and 
\BE
\label{ee}
               \epsilon= 1+ 2{{M}\over{r}}+..=e^{2\sigma}
\EE
so that Eq.(\ref{lorentz}) reduces to Eq.(\ref{basic2}).

This last argument confirms the very general
nature of Eq.(\ref{basic2}), $\sigma$ being the
Newton potential. This structure reproduces, 
in the experimentally available, slowly varying gravitational fields, 
the known classical space-time effects, and is 
in agreement with our starting point Eq.(\ref{parametric}). 
 As anticipated in the Introduction, this does not mean
that `all' possible gravitational phenomena (including those yet undiscovered)
can be described in terms of Eq.(\ref{basic2}). For instance, as
we shall comment in Sect.4, one can try to identify 
transverse vibrations in the
 scalar condensate. If these were substantially excited, 
the required space-time description
might no longer fit with the metric Eq.(\ref{basic2}). However, for the time
being, 
by restricting to the experimentally available gravitational fields, 
the known space-time effects can also be described in terms of
our medium by introducing simple
modifications of the flat Minkowski metric depending on
a scalar field $\sigma(x)$. 
This type of approach does not 
require to solve any field equations but requires to understand {\it why}
$\sigma(x)$ is the Newton potential.

\section{Long-wavelength excitations of the Higgs 
condensate}

{\bf 3.1}~~Let us now investigate a possible
 physical origin of $\sigma$. This requires
a physical framework where we can
understand its identification with the Newton
potential, i.e. a model field equation
that reduces to the Poisson equation in the 
static limit. To start with, let us first write $\sigma$ in its correct
dimensionless form. For instance, for the centrally symmetric case
the correct relation is
\BE
\label{correct}
                   \sigma(r)= {{G_N M}\over{c^2 r}} 
\EE
where 
$G_N=6.67..\cdot10^{-8}{\rm cm}^3{\rm sec}^{-2}{\rm gr}^{-1} $, 
$c=2.9979..10^{10}{\rm cm}~{\rm sec}^{-1}$, 
$M$ is in grams and $r$ in 
centimeters. Therefore, $\sigma$ is a solution of the Poisson equation
\BE
\label{simple}
\Delta \sigma= {{G_N}\over{c^2}}\rho
\EE
with 
\BE
\rho=-4\pi M \dl 
\EE

Now, since $\sigma$ is a scalar, the simplest way to
recover Eq.(\ref{simple}), is to consider a D'Alembert wave equation
with some source $S(x)$ 
that, in the static limit, becomes proportional 
to the mass density $\rho$. This leads, 
in general, to introduce a parameter $\eta$ and the 
equation 
\BE
\label{sound0}
    (\eta \Delta -{{\partial^2}\over{c^2\partial t^2}}) \sigma=-4\pi 
{ { \tilde{G}_N } \over {c^2} } S(x)
\EE
with 
\BE
\label{gtilde}
                 \tilde{G}_N= \eta G_N
\EE
The introduction of $\eta$ is useful to transform
Eq.(\ref{sound0}) into
\BE
\label{sound1}
    (\Delta -{{\partial^2}\over{c^2_s\partial t^2}}) \sigma=-4\pi
{{ {G}_N} \over{c^2}} S(x)
\EE
In fact, if
$\sigma(x)$ were associated with a medium, 
Eq.(\ref{sound1}) would describe density fluctuations propagating 
with a squared `sound velocity' 
\BE
c^2_s= \eta c^2
\EE
Let us now analyze the implications of this toy-model. Using
Eq.(\ref{sound0}), we could try to relate
the scalar field to a typical 
particle physics scale $\tilde{G}_N$. In this case, however, 
the same re-scaling $\eta$, relating
$\tilde{G}_N$ to ${G}_N$ affects the relation
between $c^2_s$ and $c^2$. Therefore, assuming a typical particle scale
for $\tilde{G}_N$, say the 
Fermi constant $G_F$, would require density fluctuations 
propagating at the fantastically high speed 
$c_s\sim 4\cdot 10^{16} c$. 
Is this conceivable ? For definiteness, we shall
explore a condensate of spinless quanta
to check whether the idea 
$c_s\sim 4\cdot 10^{16} c$ is plausible, or not.
\vskip 10 pt
{\bf 3.2}~~
Before addressing the physical properties of a scalar condensate, however, 
one premise is in order. For a spontaneously broken 
one-component $\lambda\Phi^4$ theory, i.e. 
leaving aside the Goldstone bosons,
the particle content of the broken phase
is usually represented as a single massive field, the (singlet) Higgs boson. 
Although there is no rigorous proof \cite{book}, 
the energy spectrum of the broken phase is believed to
tend to a non-zero value, $\tilde{E}({\bf{p}}) \to M_h$, 
when ${\bf{p}} \to 0$ so that the non-zero quantity $\tilde{E}(0)=M_h$ should
give rise to an exponential decay $\sim e^{-\tilde{E}(0)T}$ of the 
connected Euclidean propagator. This is equivalent to require that
the Fourier transform of the connected Euclidean propagator
tend to a finite limit,
$G(p) \to {{{\rm 1} }\over{M^2_h}}$ when the 4-momentum
$p \to 0$, with the mass squared $M^2_h$ 
being related to the quadratic shape of a semi-classical, non-convex effective 
potential $V_{\rm NC}(\phi)$ (`NC'=non-convex) at its non-trivial 
absolute minima, say $\phi=\pm v$. 

However, in a spontaneously broken phase, 
it can be shown \cite{legendre,pmu}
that $G(0)$ is a {\it two-valued} function 
that includes the case $G^{-1}(0)=0$ as in a gap-less
theory. This effect cannot be discovered in a conventional perturbative
calculation where the zero-mode of the scalar field is `frozen' at one
of the two minima of $V_{\rm NC}(\phi)$. 
To this end, one has to go beyond the simplest approximation where
the `Higgs condensate' is treated as a classical c-number field. Namely, 
one has either to perform the last functional integration
over the zero-mode of the scalar field \cite{legendre} or first re-sum
the one-particle reducible, zero-momentum tadpole graphs in the constant
background field $\phi$ \cite{pmu}, by 
taking the limit $\phi \to \pm v$ at the end of the calculation. 
The tadpole graphs, in fact, are connected to the other parts of the diagrams
through zero-momentum propagators and can be considered a manifestation of the
quantum nature of the scalar condensate.
In both cases, one finds {\it two} possible solutions
for the inverse zero-4-momentum connected
propagator at $\phi=\pm v$:
a)~$G^{-1}_a(0)=M^2_h$ and b)~$G^{-1}_b(0)=0$. 

Let us first review the results of ref.\cite{legendre}. The starting point 
is the separation of the scalar field into a constant background
and a shifted fluctuation field, namely 
\BE
\label{shift}
          \Phi(x)= \phi + h(x)
\EE
In order Eq.(\ref{shift}) to be unambiguous, 
$\phi$ denotes the spatial average in a large 4-volume $\Omega$ 
\BE
\label{average}
          \phi= {{1}\over{\Omega}}\int d^4x~ \Phi(x)
\EE
and the limit $\Omega \to \infty$ has to be taken at the end.

In this way, the full functional measure can be expressed as
\BE
\label{measure}
                \int[d\Phi(x)]...=\int^{+\infty}_{-\infty}d\phi\int[dh(x)]...
\EE
and the functional integration on the r.h.s. of Eq.(\ref{measure}) 
is over all quantum modes with 4-momentum $p\neq 0$. 

After integrating 
out all non-zero quantum modes, 
the generating functional in the presence of a space-time constant 
source $J$ is given by
\BE
\label{zetaj}
      Z(J)= \int^{+ \infty}_{-\infty} d\phi~ \exp [-\Omega
(V_{\rm NC}(\phi) - J\phi)]
\EE 
to any finite order 
in the loop expansion. Finally, by introducing the generating functional for
connected Green's functions $w(J)$ through
\BE
\label{log}
\Omega~ w(J)=\ln {{Z(J)}\over{Z(0)}}
\EE
one can compute the field expectation value
\BE
\label{phij}
\varphi(J)={{dw}\over{dJ}}
\EE
and the $p_\mu=0$ propagator
\BE
\label{GJ0}
G_J(0)={{d^2w}\over{dJ^2}}
\EE
In this framework, 
spontaneous symmetry breaking corresponds to non-zero values of 
Eq.(\ref{phij}) in the double limit $J \to \pm 0$ and 
$\Omega \to \infty$. 

Now, by denoting $\pm v$ the absolute minima of 
$V_{\rm NC}$ and 
\BE
\label{ident}
M^2_h \equiv
   {{ d^2 V_{\rm NC} }\over{d \phi^2}}
\EE 
its quadratic shape at these extrema, one usually assumes 
\BE
\label{ssb}
\lim_{\Omega \to \infty} \lim_{J \to \pm 0} \varphi(J) = \pm v
\EE
{\it and}
\BE
\label{GA}
\lim_{\Omega \to \infty} \lim_{J \to \pm 0} G_J(0)={{1}\over{M^2_h}}
\EE
In this case, the excitations in the broken phase 
would be massive particles (the conventional Higgs bosons)
whose mass $M_h$ is determined by the positive 
curvature of $V_{\rm NC}$ at its absolute minima. However, 
at $\varphi=\pm v$, 
besides the value ${{1}\over{M^2_h}}$, one also finds \cite{legendre}
\BE
\label{GB}
\lim_{\Omega \to \infty} \lim_{J \to \pm 0} G_J(0)=+\infty
\EE
a result that has no counterpart in perturbation theory. 

Analogously, 
after including the 
one-particle reducible, zero-momentum tadpole graphs, 
the formal power series for the
exact inverse zero-momentum propagator can be expressed as \cite{pmu}
\BE
\label{formal}
G^{-1}(0)= 
       \left. \frac{ d^2 V_{\rm NC}}{d \phi^2} 
\right|_{ {\hat{\phi}}=\phi(1 - \tau) } 
\EE
with
\BE
\label{tauu}
            \tau\equiv \tilde{T}(\phi^2)G(0)
\EE
where the tadpole function $\tilde{T}(\phi^2)$ vanishes at $\phi=\pm v$. 
As a result, after including tadpole graphs to all orders, 
one finds multiple solutions for the zero-4-momentum propagator 
that again can differ from (\ref{ident}) even when $\phi \to \pm v$.

In fact, Eq.(\ref{formal}) provides a regular solution
${G}^{-1}_a(0)=M^2_h$ for which
\BE
\label{taureg}
           \lim_{\phi \to \pm v} \tau = \bar{\tau}=0
\EE
and a singular solution
${G}^{-1}_b(0)=0$ such that 
\BE
\label{tausing}
           \lim_{\phi \to \pm v} \tau= \bar{\tau} \neq 0
\EE
As an example, let us consider the situation of the tree-level approximation
where $V_{\rm NC} \equiv V_{\rm tree}$ with
\BE
\label{tree}
V_{\rm tree}={{1}\over{2}} r \phi^2 + {{\lambda}\over{4!}} \phi^4 
\EE
an approximation which is 
equivalent to re-summing tree-level 
tadpole graphs to all orders (i.e. no loop diagrams).
In this case the regular solution is
${G}^{-1}_a(0)={{\lambda v^2}\over{3}}$, while the
singular solution is
\BE
\label{singular}
\lim_{\phi \to \pm v}
{G}^{-1}_b(0)=
{{\lambda v^2}\over{2}}
[ \bar{\tau}^2- 2\bar{\tau} + {{2}\over{3}}]=0
\EE
which implies limiting values 
$ \bar{\tau}=1 \pm {{1}\over{\sqrt{3}}}$.

In general, beyond the tree-approximation, 
finding the singular solution 
$G^{-1}_b(0)= 0$ at $\phi=\pm v$ is equivalent to determine that value of 
$\hat{\phi}^2\equiv v^2(1-\bar{\tau})^2$ where
${{d^2V_{\rm NC}}\over{d \phi^2}}=0$.
 For instance, in the case of the
Coleman-Weinberg effective potential
\BE
            V_{\rm NC}(\phi)= {{\lambda^2\phi^4}\over{256\pi^2}}
(\ln {{\phi^2}\over{v^2}} -{{1}\over{2}})
\EE
the required values are $\bar{\tau}=1 \pm e^{-1/3}$. In principle, 
such solutions exist in any
approximation to $V_{\rm NC}$ due to the
very general properties of the shape of a non-convex effective potential.

The b-type of solution
corresponds to processes where assorbing (or emitting)
a very small 3-momentum ${\bf{p}} \to 0$
does not cost a finite energy. This situation is well
known in a condensed medium where a very small 
3-momentum can be coherently distributed 
among a large number of elementary constituents (the hydrodynamical regime
mentioned in the Introduction). Therefore, as for $^4$He with phonons and 
rotons, in a spontaneously broken phase, there are actually 
{\it two} possible types of excitations with the same quantum numbers but
different energies when ${\bf{p}} \to 0$: 
a massive one, with $\tilde{E}_a({\bf{p}}) \to M_h$, and a gap-less one with 
$\tilde{E}_b({\bf{p}}) \to 0$. They can both propagate (and interfere) in the
broken phase. However, the gap-less excitation would dominate the 
exponential decay $\sim e^{-\tilde{E}_b({\bf{p}})T}$  of the
connected euclidean correlator for ${\bf{p}} \to 0$ so that 
the massive mode becomes unphysical in the infrared region.
Therefore, differently from the simplest perturbative indications, 
in a (one-component) spontaneously broken
$\lambda\Phi^4$ theory there would be no energy-gap associated
with the `Higgs mass'  $M_h$, as for a genuine massive single-particle 
theory where the massive spectrum 
\BE
\label{covariant}
\tilde{E}({\bf{p}}) = \sqrt{ {\bf{p}}^2 + M^2_h} 
\EE
remains true for ${\bf{p}} \to 0$. Rather, the infrared region would be
dominated by the gap-less mode. 

We observe that ref.\cite{mech}, although providing the physical mechanism
for the phion-condensation phenomenon, did not show any evidence for the
existence of the gap-less branch. This is due to the particular approximation, 
where the creation and annihilation operators 
$a_{ {\bf{p}}=0}$, $a^{\dagger}_{ {\bf{p}}=0}$
for the elementary quanta in the
${\bf{p}}=0$ mode were simply replaced by the c-number $\sqrt{N}$. 
This choice is the second-quantization
analog of `freezing' $\phi=\pm v$ without performing the functional integration
as in ref.\cite{legendre} or without first re-summing the zero-momentum tadpole 
graphs as in ref.\cite{pmu}. In this case, the only 
solution is $G^{-1}(0)=M^2_h$. 

Let us now address the physical
properties of the gap-less branch using
the standard treatment of the long-wavelength excitations in quantum 
liquids.
\vskip 10 pt
{\bf 3.3}~~
The quantum hydrodynamical picture of zero-temperature Bose systems has 
a basic universal feature: the lowest
excitations are {\it phonons}. "These are 
excited states of compression generated by 
small displacements of the elementary atoms with a resulting infinitesimal
change of their density" \cite{fey}. In other words, 
in a quantum Bose liquid"... elementary excitations with small 
momenta ${\bf{p}}$  (wavelengths large compared with distances between atoms) 
correspond to ordinary hydrodynamic sound waves, i.e. are phonons. This 
means that the energy of such quasi-particles is a linear function of their
momentum" \cite{pita2}. 
At zero temperature, they behave as 
non-interacting particles so that their energy spectrum, 
\BE
\label{etilde}
\tilde{E}({\bf{p}}) \equiv c_s |{\bf{p}}| 
\EE
( $c_s$ being the sound velocity) is virtually exact for ${\bf{p}}\to 0$.
The only restriction is due to the limitation in the ${\bf{p}}$ values for which
Eq.(\ref{etilde}) applies.
These extend up to a maximum momentum $|{\bf{p}}|_{\rm max}\equiv\delta$
which is considerably smaller than the 
inverse mean free path $R_{\rm mfp}$ for the elementary constituents

Therefore, in a zero-temperature scalar condensate, and 
in the limit ${\bf{p}} \to 0$, quantum hydrodynamics 
 predicts long-wavelength excitations
to be density fluctuations. In this approach, 
the phion condensate undergoes 
small oscillations where the local density 
$ n({\bf{r}})$ is very close to its equilibrium value $n_o$, i.e.
$|n({\bf{r}})-n_o|\ll n_o$. Also, 
the current flow is such that $|n{\bf{v}}| \ll n_o c_s$ and
the velocity field ${\bf{v}}$ describes a
potential flow
\BE
\label{flow}
                          {\bf{v}}=\nabla \Phi
\EE
so that one can replace $\nabla({ n {\bf{v}} }) \sim n_o \Delta \Phi$, 
obtaining the `Poisson regime'.

Thus, we understand the gap-less
mode discovered in refs.\cite{legendre,pmu} as being due to the propagation of 
phonons in the phion condensate and we can start to
figure out the possible values of $c_s$.
To this end, we first observe that the idea of a
`quantum aether' was considered by Dirac \cite{dirac}. 
In his approach, there is a time-like
vector field $v_\mu$, with $v_\mu v^\mu=1$, that, 
in a classical picture, would correspond to an
aether velocity field. 
For a quantum theory, rather than considering operatorial relations, 
one may introduce \cite{dirac}
a wave function for the aether and require its modulus to be 
independent of $v_\mu$, as with plane-waves 
in quantum mechanics. Let us exploit some consequence of this idea and
consider the usual Klein-Gordon equation for particles of physical
mass $m^2_\Phi >0$ 
\BE
\label{klein}
     ({{\partial^2}\over{\partial t^2}} - \Delta) \Phi=-m^2_\Phi \Phi
\EE
We shall restrict to solutions of the form $\Phi=\Phi(w)$ 
where $w=v_\mu x^\mu$ is the Lorentz-invariant variable defined at each
space-time point in a quantum aether with a given constant 
$v_\mu\equiv (v_o, {\bf{v}})$.
Periodic solutions have the form $\hat{\Phi}=a~ cos(m_\Phi w)+b~sin(m_\Phi w)$ 
and would usually be interpreted as particles with 4-momentum 
$p_\mu=m_\Phi v_\mu$ and mass-shell condition $p^2=m^2_\Phi$. 

However, let us slightly change perspective and interpret the same
Klein-Gordon equation in a statistical sense, i.e. as the equation governing 
the collective wave-function of a 
`Madelung fluid' \cite{bohm} of phions moving with velocity ${\bf{v}}$.
 In this case, the {\it same} periodic solutions, satisfying the wave-equation
\BE
\label{sound2}
    (c^2_s \Delta -{{\partial^2}\over{\partial t^2}}) \hat{\Phi}(w)=0
\EE
with
\BE
c^2_s\equiv 1 +  {{1}\over{ |{\bf{v}}|^2 }}~~,
\EE
 would naturally be interpreted as {\it density waves}, 
in agreement with the phonon energy spectrum Eq.(\ref{etilde})
expected on the base of Landau's hydrodynamical picture. 

Notice that the phonon-like excitations of the phion condensate 
propagate with a superluminal velocity $c_s >1$. Therefore, near the
aether rest-frame, i.e. where $|v_o|\sim 1 $ and $|  { \bf {v }} |^2 \to 0 $,
these waves would be seen to
propagate with a very large speed $c_s \to \infty$. In addition,  
with the same space-time units, one can introduce a space-like vector 
$s_\mu$ such that $s^\mu s_\mu=-1$ and $s_\mu v^\mu=0$ as for a genuine 
vacuum angular momentum. By considering the 
same Klein-Gordon equation (\ref{klein}) for a 
function $\chi=\chi(\xi)$, $\xi$ being the other Lorentz-invariant combination
$\xi=s_\mu x^\mu$, the equivalent waves associated with 
$\chi$ propagate with a subluminal velocity $c_s <1$. 

Obviously, 
analogous results hold in a reversed form. For instance, let us start from any
superluminal wave equation with $c^2_s\equiv 1+ \beta^2> 1$ 
\BE
(c^2_s \Delta-
{{\partial^2}\over{\partial t^2}})\varphi=0
\EE
and consider its plane wave solutions $\varphi=\varphi(w)$, with 
$w= \sqrt{1 + {{1}\over{\beta^2}}}t - {{1}\over{\beta}} {\bf{n}}\cdot {\bf{r}}$, 
depending on an arbitrary spatial direction 
${\bf{n}}$ with ${\bf{n}}\cdot{\bf{n}}=1$. 
The periodic solutions $\hat{\varphi}(\tau)$, such that
${{d^2 \hat{\varphi} } \over{d\tau^2}}=-\hat{\varphi}$, depend on a 
dimensionless parameter 
$\tau\equiv \alpha w$, $\alpha$ being a physical mass scale introduced 
to define the periodicity condition in some units. 
They are solutions of a Klein-Gordon equation 
\BE
\label{lorentz2}
 ({{\partial^2}\over{\partial t^2}} - \Delta) \hat{\varphi}=-\alpha^2 
\hat{\varphi}
\EE
with a physical mass-squared $\alpha^2 >0$. In this case, the 
Lorentz-covariant equation emerges when the dependence on the hidden 
mass scale $\alpha$ is made explicit.

The same idea that density fluctuations
in a Higgs condensate propagate with a velocity 
$c_s$ which is {\it infinitely} 
larger than the speed of light
is also suggested by a semi-classical
argument due to Stevenson \cite{seminar} that we shall briefly report.
Stevenson's argument starts from a perfect-fluid treatment of the Higgs
condensate in its rest frame. The perfect-fluid approximation is frequently
used in the literature, for instance in cosmology. If there is a Higgs 
condensate around, this approximation may also apply to it.
Now, in this approximation, 
energy-momentum conservation is equivalent to wave propagation with a
squared velocity given by 
\BE
\label{cs}
        c^2_s= c^2 ({{\partial {\cal P}}\over{\partial {\cal E} }})
\EE
where ${\cal P}$ is the pressure and ${\cal E}$ the
energy density. Introducing the condensate density $n$, and using
the energy-pressure relation
\BE
\label{zerot}
 {\cal P}= -{\cal E} + n {{\partial {\cal E} }\over{\partial n}}
\EE
we obtain
\BE
\label{nn1}
c^2_s=
c^2({{\partial {\cal P}}\over{\partial n}})
({{\partial {\cal E} }\over{\partial n}})^{-1}=
c^2  (n{{ \partial^2 {\cal E} }\over{\partial n^2}})
({{\partial {\cal E} }\over{\partial n}})^{-1} 
\EE
For a non-relativistic Bose condensate of neutral particles with mass
$m$ and scattering length $a$, where ${{na\hbar^2}\over{m^2c^2}} \ll 1$
(in this case we also explicitely introduce $\hbar$) one finds
\BE
\label{nonrel}
{\cal E}= n m c^2 + n^2 {{2\pi a \hbar^2}\over{m}} 
\EE
so that the sound velocity is 
\BE
\label{nn2}
c^2_s={{4\pi n a \hbar^2}\over{m^2}}
\EE
Eq.(\ref{nn2}) agrees with the sound velocity obtained from the 
{\it microscopic}
\cite{huang} derivation of the Bogolubov spectrum in a 
dilute hard-sphere Bose gas. 

 On the other hand, when describing the occurrence of spontaneous symmetry 
breaking, where the scalar quanta are phions with mass $m_\Phi$, 
there are now additional terms
in Eq.(\ref{nonrel}). These are such that the scalar condensate 
is spontaneously generated from the `empty' vacuum where $n=0$ for
that particular equilibrium phion density where \cite{mech}
\BE
\label{vacuum}
{{\partial {\cal E} }\over{\partial n}} = 0
\EE
Therefore, in this approximation, approaching the equilibrium density
one finds
\BE
\label{infinity}
           c^2_s \to \infty
\EE
implying, again, that long-wavelength
density fluctuations would propagate instantaneously in the 
spontaneously broken vacuum. 

As Stevenson points out \cite{seminar}, 
Eq.(\ref{infinity}) neglects all possible corrections to the perfect-fluid
approximation, just as Eq.(\ref{klein}) neglects the effect of a weak phion
self-coupling $\lambda$.
 These phion-phion interactions introduce collisional effects associated
with a finite mean free path $R_{\rm mfp}\sim {{1}\over{na^2}} $, $n$ being 
the phion number density and $a$ their S-wave scattering length 
\cite{kinetics}. Due to its finite value, 
density waves will propagate at a large but finite speed $c_s$ such that 
$c_s \to \infty$ when $\lambda\to 0$.  Also their
wavelengths will be larger than
$R_{\rm mfp}$ \cite{seminar} in agreement with the hydrodynamical 
restriction of the energy spectrum Eq.(\ref{etilde}) to momenta 
$|{\bf{p}}| < R^{-1}_{\rm mfp}$.

Finally, the result $c_s \to \infty$ is also suggested by a third argument, 
if one takes into account the approximate nature of locality 
in cutoff-dependent quantum field theories. In this picture, the elementary
quanta are treated as `hard spheres', as for the molecules of ordinary
matter. Thus, the notion of the vacuum as a `condensate' acquires an
intuitive physical meaning. For the same reason, however, the simple idea
that deviations from Lorentz-covariance take only place at the cutoff scale
may be incorrect. In fact, 
a hard-sphere radius is known, from the origin of
Special Relativity \cite{born}, to imply a superluminal propagation
within the sphere boundary. Now, in the perturbative empty vacuum state 
(with no
condensed quanta) such superluminal propagation is restricted to very short
wavelengths, smaller than the inverse ultraviolet cutoff. However, in the
condensed vacuum, the hard spheres can `touch' each other so that the 
actual propagation of 
density fluctuations in a hard-sphere system
might take place at a superluminal speed. This is very close to our previous
`dual' picture of the Klein-Gordon equation. Although the individual 
particles are limited to a time-like motion, there are real physical 
collective excitations, in the form of density waves \cite{vigier}
 that propagate at a superluminal speed. This provides a definite model 
\cite{vigier} for the non-local nature of the quantum potential \cite{bohm}
and, as such, a possible answer to the delicate questions raised by 
Bell's inequality. 

In this sense, hard-sphere condensation, as a model of the broken-symmetry
vacuum in a cutoff theory, leads to what 
Volovik calls {\it reentrant violations of special relativity in the
low-energy corner} \cite{reentrant}. This produces
a simple physical picture in terms of the two different solutions 
\cite{legendre,pmu} for the zero-momentum propagator. 
In fact, let us consider an infinite, isotropical 
scalar condensate where both translational and rotational invariance is
preserved. In this case, 
the possible {\it reentrant} violations will extend 
over a small shell of momenta, say $|{\bf{p}}| < \delta$, where the 
condensate excitation spectrum $\tilde{E}=\tilde{E}(|{\bf{p}}|)$ 
deviates from a Lorentz-covariant form. 
However, full Lorentz covariance has to be re-established 
in the local limit. Therefore, for a large but finite 
ultraviolet cutoff $\Lambda$, the scale $\delta$ is naturally
infinitesimal in units of the scale associated with the Lorentz-covariant 
part of the energy spectrum, say $M_h$. 
By introducing dimensionless quantities, this means 
$\epsilon\equiv {{\delta}\over{M_h}} \to 0$ when
$t \equiv {{\Lambda}\over{M_h}} \to \infty$ so that
the continuum limit can equivalently be defined either as $t \to \infty$ 
{\it or}
$\epsilon \to 0$. Notice that, formally, 
${\cal O}({{\delta}\over{M_h}})$ vacuum-dependent corrections would
represent ${\cal O}({{M_h}\over{\Lambda}})$ effects which are always
neglected when discussing \cite{nielsen} how
Lorentz-covariance emerges at scales much smaller than the ultraviolet 
cutoff. Therefore, 
although Lorentz-covariance is formally recovered in the local
limit one finds, for large but finite $\Lambda$, infinitesimal deviations
in an infinitesimal region of momenta. 

Now, for our phion condensate, the 
quantum hydrodynamical analysis predicts the nature of 
the energy spectrum for ${\bf{p}}\to 0$. This has 
a very general nature, independently of any detail of the
theory at the scale $\Lambda$, and corresponds to
phonons as in Eq.(\ref{etilde}) 
propagating with a sound velocity $c_s$, up to 
momenta $|{\bf{p}}|\sim \delta$ that start to be comparable with the inverse
mean free path $R^{-1}_{\rm mfp}$. There, we expect a transition to a 
Lorentz-covariant single-particle spectrum as in Eq.(\ref{covariant}), 
 so that 
\BE
\label{beyond}
\sqrt{ \delta^2 + M^2_h} \sim c_s \delta
\EE
or
\BE
\label{newmatch}
             c_s\sim {{M_h}\over{\delta}}={\cal O}({{1}\over{\epsilon}})
\EE
This confirms that $c_s$, potentially, is an infinitely                    
large quantity that diverges in the continuum limit. In fact, for a fixed 
mass scale $M_h$, the limit
$t \to \infty$ can be replaced, equivalently, by $\epsilon \to 0$ or
$c_s \to \infty$. 

Due to this result, 
the massive branch will become predominant at momenta that are 
higher than $\delta$. In fact, in this region, 
the collective excitations become unphysical
since $c_s|{\bf{p}}|$ is now much larger than 
$\sqrt{ {\bf{p}}^2 +M^2_h}$. Once more, 
 this shows that the phion condensate, exhibiting a double-valued 
propagator \cite{legendre,pmu} and thus allowing for the propagation of
two different types of excitations, is very close to superfluid 
$^4$He. There, the existence of two types of
excitations  was first deduced theoretically by Landau
on the base of very general arguments \cite{hydro}. According to that original
idea, there would be phonons with energy
$E_{\rm ph}({\bf{p}}) = v_s |{\bf{p}}|$ and
rotons with energy
$E_{\rm rot}({\bf{p}})= \Delta + {{ {\bf{p}}^2 }\over{2\mu}}$. 
Only {\it later}, it was experimentally discovered that
there is a {\it single} energy spectrum $E({\bf{p}})$ which is
made up by a continuous matching of these two
different parts. This unique spectrum agrees with the phonon branch
for ${\bf{p}} \to 0$ and agrees with the roton branch 
at higher momenta.

We observe that, in the continuum limit ${{\Lambda}\over{M_h}} \to \infty$,
superluminal wave propagation is restricted to the region 
$|{\bf{p}}| < \delta$ with ${{\delta}\over{M_h}} \to 0$. Therefore, 
in a strict local limit, it would be impossible to use these waves to 
form a sharp wave front and transfer informations with
violations of causality \cite{seminar}. These require values
${{d\tilde{E}}\over{ d|{\bf{p}}| }} >1 $ at large $|{\bf{p}}|$ 
that cannot occur due to the change of the energy spectrum from 
Eq.(\ref{etilde}) to Eq.(\ref{covariant}). On the other hand, for finite 
$\Lambda$, where $\delta$ is a finite scale, the same conclusion is not so
obvious. In particular, the {\it reentrant} nature of the deviations 
from Lorentz covariance at small $|{\bf{p}}|$ is related to
the genuine superluminal nature of the Fourier spectrum
for $|{\bf{p}}| > \Lambda$ in cutoff theories. Therefore, there might be some 
differences with respect to other analyses \cite{peters} where the condition 
${{d\tilde{E}}\over{ d|{\bf{p}}| }} >1 $ is shown 
not to be in conflict with causality. The nature 
of the problem, by itself, would deserve a dedicated 
effort. Here, we shall limit ourselves to these remarks and
comment in the conclusions on possible
superluminal effects in
General Relativity, with and without a cosmological constant.

As anticipated, the limit 
$c_s \to \infty$ tries to simulate an exact
Lorentz-covariant theory where 
the energy spectrum maintains its massive form 
down to ${\bf{p}} = 0$. Yet, 
this is not entirely true due to the subtleties
associated with treating 
the zero-measure set ${\bf{p}}=0$. This set, in fact, belongs to the range
of Eq.(\ref{etilde}) and therefore
the right $c_s = \infty$ limit is always
$\tilde{E}({\bf{p}}=0)=0$ and not 
$\tilde{E}({\bf{p}}=0)=M_h$. 
For this reason, the correct procedure is to 
include both branches of the spectrum in the representation of 
the fluctuation field 
\BE
h(x)=\Phi(x) -\langle\Phi\rangle
\EE
By expanding in eigenmodes of the momentum, 
it may be convenient, however, to separate out 
the component of the fluctuation associated with the
long-wavelength modes Eq.(\ref{etilde}), say $\tilde{h}(x)$, 
 from the more conventional massive part of Eq.(\ref{covariant}). 
The observable effects due to the excitation of
 $\tilde{h}$ depend crucially 
 on the value of $c_s$ and will be discussed below.
\vskip 10 pt
{\bf 3.4}~~
Let us ignore, for the moment, all previous indications for a very large
$c_s$ and just explore the phenomenological implications
of long-wavelength modes in the spectrum as in Eq.(\ref{etilde}).
Whatever the value of $c_s$, 
these dominate the infrared region so that
a general yukawa coupling of the Higgs field to fermions will give
rise to a long-range {\it attractive} potential
between any pair of fermion masses $m_i$ amd $m_j$ 
\BE
\label{Newton}
            U_{\infty}(r)= - {{1}\over{4\pi c^2_s
\langle \Phi \rangle^2 }}{{m_im_j}\over{r}}
\EE
The above result would have a considerable impact
for the Standard Model if we take the usual value $\langle\Phi\rangle$
related to the Fermi constant. Unless $c_s$ be an extremely 
large number (in units of $c$) one is faced with strong
long-range forces coupled to the inertial masses of the known elementary 
fermions that have never been observed. 
 Just to have an idea, for $c_s = c$ 
the long-range interaction between two electrons in
Eq.(\ref{Newton}) is ${\cal O}(10^{33})$ larger than their purely
gravitational attraction. 
On the other hand, invoking a phenomenologically viable strength,  
as if $c_s \langle\Phi\rangle$ were
of the order of the Planck scale, is equivalent to re-obtain 
nearly instantaneous interactions transmitted by the scalar condensate as 
in Eq.(\ref{infinity}). 

Independently of phenomenology, a direct proportionality relation between
$c_s \langle\Phi\rangle$ and $M_{\rm Planck}$ 
is also natural noticing that
for $c_s \to \infty$ 
the energy-spectrum becomes Lorentz-covariant (with the exception
of ${\bf{p}}=0$). Therefore, 
in a picture where the `true' dynamical origin
of gravity is searched into long-wavelength deviations from exact
Lorentz-covariance, it would be 
natural to relate the limit of a vanishing gravitational strength, 
$M_{\rm Planck}\to \infty$, to the limit of
an exact Lorentz-covariant spectrum, 
$c_s \to \infty$.

Returning to more phenomenological aspects, we observe that the potential 
in Eq.(\ref{Newton}) can also be derived as a static limit 
from the effective lagrangian
\BE
\label{lagrangian1}
 {\cal L}_{\rm eff} (\tilde{h})=  {{1}\over{2}} \tilde{h}
       [ \eta \Delta - {{\partial^2 } \over{c^2\partial t^2}}] \tilde{h}
      -  {{ \tilde{h} } \over{ \langle \Phi\rangle}}
        \sum_f m_f c^2 \bar{\psi}_f \psi_f
\EE
where the free part takes into account the peculiar nature of the energy
spectrum Eq.(\ref{etilde}) and we have defined 
\BE
\label{etacs}
c^2_s\equiv\eta c^2
\EE
Eq.(\ref{lagrangian1}) is useful to represent the effects of $\tilde{h} $ 
over macroscopic scales as required by its
long-range nature. The key-ingredient is the replacement of
        $mc^2 \bar{\psi} \psi$ with 
$T^{\mu}_{\mu}(x)$, the trace of the energy-momentum tensor of ordinary matter,
a result embodied into the well known relation
\BE
        \langle f| T^{\mu}_{\mu}| f \rangle=
         m_fc^2 \bar{\psi}_f \psi_f
\EE
If we neglect quantum fluctuations, 
this relation allows for an intuitive 
transition from the quantum to the classical
theory. In fact, by introducing 
a wave-packet corresponding to a particle of
momentum ${\bf{p}}$ and normalization 
$\int d^3{\bf{x}} \bar{\psi} \psi={{mc^2}\over{E({\bf{p}})} }$
we obtain 
\BE
\label{replace}
        - m c^2
\int d^4x \bar{\psi} \psi~ =~-mc \int ds
\EE
where $ds=cdt\sqrt{1- {{ {\bf{u}}^2}\over{c^2 }} }$ 
denotes the infinitesimal element of proper time for a classical
particle with 3-velocity ${\bf{u}}$. Therefore, using the relation
\BE
         \sum_n m_n c_n \int ds_n= \int d^4x T^{\mu}_{\mu}(x)
\EE
where
\BE
         T^{\mu}_{\mu}(x) \equiv \sum_n 
{ { E^2_n -   c^2{\bf{p}}_n  \cdot  {\bf{p}}_n  } \over{E_n}} 
\delta^3 ( {\bf{x}} - {\bf{x}}_n(t) ) 
\EE
Eq.(\ref{lagrangian1}) is replaced by
\BE
\label{lagrangian}
 {\cal L}_{\rm eff} ({\sigma})=  
{{1}\over{2}}\tilde{h}
       [ \eta \Delta - {{\partial^2 } \over{c^2\partial t^2}}] \tilde{h}
      - {{ \tilde{h} } \over{ \langle \Phi\rangle}}  T^{\mu}_{\mu} 
\EE
with the equation of motion
\BE
\label{step1}
       [ \eta  \Delta - {{\partial^2 } \over{c^2\partial t^2}}] 
{{ \tilde{h} } \over{ \langle \Phi\rangle}}=
         { { T^{\mu}_{\mu} }\over{ \langle \Phi \rangle^2}}
\EE
Finally, 
by comparing with 
Eqs.(\ref{sound0}) and (\ref{sound1}), 
we find 
\BE
\label{fund0}
{{ \tilde{h}(x) } \over{ \langle \Phi\rangle}}
= - \sigma(x)
\EE
\BE
\label{fund1}
{{4\pi \tilde{G}_N }\over{c^2}} ={{1}\over{\langle \Phi \rangle^2} }\equiv G_F
\EE
and 
\BE
\label{fund2}      
S(x)=  T^{\mu}_{\mu} (x)
\EE
In fact, for the
large values of $\eta $ that are suggested by the properties of the vacuum, 
the effects of
${\tilde{h}}$ have practically no retardation effects. 
In this limit, Eq.(\ref{step1}) reduces to an instantaneous interaction
\BE
\label{step2}
          \Delta 
{{ \tilde{h} } \over{ \langle \Phi\rangle}}=
{{ T^{\mu}_{\mu} }\over{\eta  \langle \Phi \rangle^2 }} 
\EE
of vanishingly small strength when $\eta \to \infty$.
Finally for very slow motions, when the trace of the energy-momentum tensor 
becomes proportional to
the mass density Eq.(\ref{rho}), 
one re-obtains, formally, the Poisson equation 
with the Newton constant $G_N$ expressed as 
\BE
\label{fund3}
         G_N= {{G_F c^2}\over{4\pi \eta }}
\EE
This fixes the values of $\eta$ giving
 $c_s=\sqrt{\eta}c\sim 4\cdot 10^{16}c $, as anticipated.

Assuming the above value for $c_s$, the relation (\ref{beyond})
$\delta\sim {{M_hc^2}\over{c_s}}$, and
depending on the actual value chosen for 
$M_hc^2={\cal O}(\langle\Phi\rangle)$, one finds a
length scale $R_{\rm mfp}=\delta^{-1}$ in the millimeter
range ($\sim8$ millimeters for $M_hc^2=1$ TeV). 
In this framework, 
the tight infrared-ultraviolet connection embodied in the relation 
$\delta \sim {{\langle\Phi\rangle^2}\over {M_{\rm Planck}}}$ is
 formally identical to that occurring 
in models \cite{dimo} with extra space-time dimensions 
compactified at a size $R_{\rm c}=R_{\rm mfp}$.

For $r\sim R_{\rm mfp} = \delta^{-1}$ the 
interparticle potential is not a simple $1/r$, as for asymptotic distances, 
but has to be computed from the 
Fourier transform of the $h-$field propagator
\BE
D(r)=
\int {{d^3 {\bf{p}} }\over{(2\pi)^3 }}
 {{e^{ i {\bf{p}}\cdot {\bf{r}} } }\over{ 
\tilde{E}^2({\bf{p}}) }}
\EE
and depends on the detailed form of the spectrum that interpolates between
Eqs.(\ref{etilde}) and (\ref{covariant}). In this picture, the millimeter
range marks the typicale scale of `fifth-force' experiments.

Notice that a similar picture would have been obtained by skipping 
subsects. (3.2)-(3.4) and simply
identifying $\sigma(x)$ with an hypothetical 
`dilaton' field $\tilde{d}$. This would be coupled to the trace
of the energy-momentum tensor and replace our 
long-wavelength fluctuation field
${{ \tilde{h} } \over{ \langle \Phi\rangle}}$.
However, in this approach, the Newton constant
has to be introduced from scratch as a fundamental scale that fixes the
dilaton coupling. Alternatively, if
the dilaton coupling is fixed by the Fermi scale $G_F$
one has to solve the `hierarchy problem', i.e. to replace our naturally,
infinitely large $c_s$ with some equivalent mechanism to explain the 
difference between $G_F$ and $G_N$.

\section{Summary and outlook}

There are two different aspects of our analysis. On one hand, one can 
produce a simple picture of gravity in terms of
a medium characterized by a scalar field $\sigma(x)$ that, for slowly 
varying fields, is known experimentally 
to coincide with the Newton potential.

On the other hand, independently of gravity, one is faced with
the existence of a (non-Goldstone) gap-less mode of the Higgs field in the 
broken-symmetry phase. 
In fact, the simple perturbative idea of a purely massive singlet Higgs boson
field depends on treating
the scalar condensate as a classical 
c-number field. Beyond this level of approximation, 
$G^{-1}(p=0)$ is a two-valued function 
\cite{legendre,pmu} that includes the value $G^{-1}(p=0)=0$, as in a 
gap-less theory. Exploiting the possible implications of this result
should be very natural. After all, the Higgs field was introduced 
to obtain a consistent quantum theory and 
it would be really paradoxical to conclude that the Standard Model can only
work provided we treate its vacuum state as a purely classical c-number field.
At the same time, if the Higgs vacuum is considered a real superfluid medium, 
made up of physical
spinless quanta, it might even be {\it postulated} \cite{consoli}
that there are density fluctuations with an energy spectrum 
$\tilde{E}({\bf{p}})\sim c_s |{\bf{p}}|$ when 
${\bf{p}} \to 0$. This would give rise to an attractive $1/r$ 
potential among all particles coupled to the (singlet) Higgs field that
has to be understood. 
Alternatively, one can study the effect of
Bose condensation on the forces among bodies sitting in a 
`$\lambda\Phi^4$ ambiance'
\cite{ferrer}. Below the transition temperature, i.e. in the broken-symmetry 
phase, the range of the forces becomes infinite and one finds
a $1/r$ potential. Again, this shows that the condensate energy spectrum 
cannot be a pure $\sqrt{ {\bf{p}}^2 + M^2_h}$ down to ${\bf{p}}=0$ since,
 in this case, there would be no long-range force.

Just for this reason, looking for 
the physical origin of $\sigma(x)$, we propose a natural candidate:
the collective density fluctuations of the
scalar condensate. These propagate as longitudinal waves, 
starting at wavelengths that are larger than the mean free path
$R_{\rm mfp}$ for the elementary phions and phenomenology requires 
$R_{\rm mfp}$ to be a length scale in the millimeter range. For this
reason, there will be no variation of the 
gravitational potential between two points whose distance is smaller than
$R_{\rm mfp}$ since the collective oscillations of the condensate have
larger wavelengths. These average over distances that are much
larger than the atomic size so that all quantum 
interference effects disappear, in agreement with the point of view
expressed in the Introduction. An exception 
is represented by those particular experiments where the coherence of the
wave-functions can be maintained over distances where $\sigma(x)$ can 
vary appreciably, as for the 
neutron diffraction experiments in the earth's gravitational field
\cite{cow}. In such cases, an acceleration transformation to a suitable
freely falling frame, to eliminate the effects of the direct coupling of 
$\sigma(x)$ to all particles, 
will introduce mass-dependent phases for the various
wave-functions. Therefore, in principle, through
quantum interference experiments, one might distinguish between the two frames.

The vastly superluminal value of the `sound velocity' $c_s=\sqrt{\eta}c$, with
$\sqrt{\eta} = 4\cdot 10^{16} $, needed to relate the Newton constant 
$G_N$ to the Fermi constant $G_F$, is the only free parameter of our analysis.
However, its magnitude 
is not totally unexpected on the base of the properties of the Higgs
condensate. Indeed, it is consistent with {\it three} different
arguments, all pointing toward $c_s\to \infty$, that would motivate
the mysterious, nearly instantaneous nature of Newtonian gravity 
\cite{tom} at the base of its 
traditional interpretation as an `action at distance'. This picture provides
a simple physical solution of the `hierarchy problem' where the large value
of the dimensionless ratio ($ \delta \sim R^{-1}_{\rm mfp}$) 
\BE
\label{hiera}
\sqrt{\eta} \sim {{M_h}\over{\delta}}\sim {{M_{\rm Planck}}\over{M_h}}=
{\cal O}( 10^{16})
\EE
derives from the
Lorentz-covariance of the energy spectrum in the cutoff theory 
down to $|{\bf{p}}|\sim \delta$ (see Eq.(\ref{beyond})), with 
${{\delta}\over{M_h}} \to 0$ in the continuum limit. 

It is interesting 
that the various quantities entering Eq.(\ref{hiera}) admit 
a simple interpretation in terms of the two
basic quantities of the phion condensate: the phion density $n$ and the phion
scattering length $a$. In terms of these quantities we find \cite{mech}
$M^2_h \sim na$ and $R_{\rm mfp} \sim {{1}\over{na^2}}$ \cite{seminar}.
Therefore, using our results from subsect.3.4, we find 
\BE
        M_h R_{\rm mfp} \sim {{1}\over{ \sqrt{ na^3} }} = 
{\cal O}(10^{16})
\EE
with a scattering length of the order of the Planck length 
\BE
\label{dilu}
       a \sim na^3 R_{\rm mfp} \sim {\cal O} (10 ^{-33})~cm
\EE
since the 
`diluteness factor' $na^3 \sim {\cal O}(10^{-33})$ is extremely small. 

Before concluding, we shall 
mention some points that deserve further
study:

~~~i) the idea of the Higgs condensate as a real superfluid medium suggests
alternative scenarios for the Higgs boson production, 
in addition to the standard mechanisms
(e.g through quark pair, W-pair,..annihilation).  
For instance, considering high-energy cosmic ray proton-proton collisions, 
the whole energy content of the initial state $\sqrt{s}\sim \sqrt{2 m_p E}$ might
be used for a `local heating' of the vacuum \cite{mishra}. A substantial
heat release might drastically 
excite the superfluid vacuum and give rise to dissipative 
processes, analogously to the shock waves produced
by a moving body with 
current density $|{\bf{J}}({\bf{r}})|> c_s n$. If we identify
$M_hc^2 \sim c_s \delta$ as the relevant energy scale,
we would tentatively conclude that the local-heating mechanism 
might become efficient for center of mass energies $\sqrt{s}> M_hc^2$ or
for cosmic ray 
energies $E> 5x^2\cdot 10^{14}$ eV where $x$ is the value of $M_hc^2$ 
in TeV. In this case, a sizeable fraction of the 
primary flux might be used to excite the vacuum, rather than to produce
leading hadrons with the associated electromagnetic 
showers. Therefore, on the ground, one would count less events, as if
the primary flux would have been reduced, and one might speculate on 
alternative interpretations of the famous `knee' \cite{kulikov} in the
cosmic ray spectrum whose precise position and physical origin are still
unclear \cite{wiebel}. For instance, early investigations \cite{akeno}
showed a rather sharp peak at 
$E\sim 5\cdot 10^{15}$ eV whereas newer measurements \cite{tibet,norikura}
favour a more gradual steepening starting at $E\sim 1-2\cdot 10^{15}$ eV. 
At the same time, there may be some
inconsistencies in the standard astrophysical interpretations 
\cite{wiebel,niko3} that could motivate the idea \cite{nikolski} 
that some
new state ("...strongly interacting bosons with masses $>$400 GeV/$c^2$.." 
\cite{niko2}) is indeed produced in the primary collision. 

~~~ii)  although all metrics discussed by Tupper
agree in the weak-field limit \cite{tupper}, we have explained why
Yilmaz's metric would play a special role: in this case, the metric 
depends on the scalar field $\sigma(x)$ in a parametric form so that Einstein's
field equations are actually 
algebraic identities. This is a consistency requirement
in a theory where gravity is an effective interaction induced by the vacuum 
of some underlying quantum field theory. However, Yilmaz's theory differs
from General Relativity for strong fields where
the Schwarzschild singularity 
${{1}\over{1-2\sigma}}$ 
becomes an exponential form $e^{2\sigma}$. For a many-body gravitational 
system, the 
unique factorization properties of the Yilmaz metric, 
$e^{\sum_i {{M_i}\over{r_i}} }=e^{{{M_1}\over{r_1}}} e^{{{M_2}\over{r_2}}}..$,
provide an alternative 
explanation for the controversial huge 
quasar red-shifts, a large part of which could be interpreted
as being of gravitational (rather than cosmological) origin \cite{clapp}.
 This might represent a
`fifth' test of gravity outside of  the weak-field
regime.

~~~iii) in our approach, $T^\mu_\mu$ is the source
of the long-wavelength density fluctuations
of the Higgs condensate and, as such, of Newtonian gravity. This is
a specific consequence of the coupling to a 
scalar field. In this sense, the Higgs vacuum provides the natural
framework for a common dynamical origin of inertia and gravity that would
both disappear without the scalar condensate, i.e. in the limit 
$\langle\Phi\rangle \to 0$. It also provides 
an explicit realization \cite{consoli}
of the {\it Mach's Principle} where the vastly superluminal value
of $c_s$ might
represent the non-local element to understand the apparent a-causal nature
of the inertial reactions in an accelerated frame \cite{queery}. 

~~~iv) as discussed by Dicke \cite{varenna}, when averaged over 
sufficiently long times (e.g. with respect to the atomic times), 
by the virial theorem \cite{champ}, the integral of $T^\mu_\mu$ 
represents the total energy of a 
bound system,  i.e. includes the binding energy.
Therefore, for microscopic systems whose components have 
large $v^2/c^2$ but very short periods, this definition 
becomes equivalent to the rest energy. On the other
hand, for macroscopic systems, that have long periods but small $v^2/c^2$, 
there should be no observable
differences from the mass density. A possible exception might be associated
with the overall motion
of our galaxy, when assuming for this velocity the sizeable value
${{v}\over{c}} \sim 10^{-3}$ \cite{varenna}. This velocity affects all known
classical sources of gravity and should represent an
overall re-definition of their mass. However, 
small differences may eventually be found
by precise observations of the GPS satellites.
These are placed on
nearly circular orbits of radius $r_{\rm GPS}\sim 26600$ Km and periods
$T_{\rm GPS}\sim 11$ hours and 58 minutes \cite{tomgps}. If the trace of the
energy-momentum tensor is the true source of gravity, 
the GPS Keplerian `invariant' $T\cdot r^{-3/2}$ should change by about one
part over $10^7$ when the earth's velocity  
is parallel or antiparallel to the galactic velocity. Although small, this 
effect could be observable in view of the spectacular accuracy of the 
GPS system.

~~~v) one should not conclude that the (nearly)
instantaneous nature of 
Newtonian gravity, in our picture, is in contradiction with 
binary pulsar data showing that gravity is due to "... massless spin-2 
gravitons propagating at the speed of light". In fact, the observed slowing
down of the binary systems, when interpreted in terms of a quadrupole 
gravitational radiation, could at best be used to predict {\it deviations}
from exact Keplerian orbits and not the Keplerian orbits themselves. 
This means that, in principle, the slowing down of binary pulsars and 
the physical mechanism for Newtonian gravity represent separate issues whose
possible conceptual unification depends on the theoretical framework. In
General Relativity, unification consists in introducing the same entity, the 
graviton field $h_{\mu \nu}$, with $g_{\mu \nu}=\eta_{\mu \nu} + h_{\mu \nu}$, 
where $h_{44}= {{2M}\over{r}}$ is used for the Newton potential and 
the transverse components account for gravitational radiation. However, 
condensed media are full of examples where longitudinal 
and transverse excitations do not propagate with the same velocity. 
In addition, it is now well known \cite{recami} that 
the standard D'Alembert wave equation 
can exhibit superluminal solutions ( there are even some {\it experimental}
evidences \cite{mugnai}). This can also
be checked following our own simple argument. Let us define 
$\eta=z-ut$ and $r=\sqrt{ {{ u^2-c^2}\over{c^2}}} \sqrt {x^2 + y^2}$ where
$u^2 > c^2$. Any function $\psi=\psi(\eta,r)$ describes `cylindrical waves' 
propagating along the z-axis with a velocity $u>c$ since $\psi$ has 
the same value for $z-ut= {\rm const}$ and 
$x^2+y^2={\rm const}$. However, one can find special types of
$\psi$'s that are
solutions of the D'Alembert wave equation, i.e. for which
\BE
\label{para}
        ({{ \partial^2}\over{\partial x^2}} +              
        {{ \partial^2}\over{\partial y^2}} +              
        {{ \partial^2}\over{\partial z^2}} -              
        {{ \partial^2}\over{c^2\partial t^2}})\psi=0               
\EE
In fact, with our choice of $\psi$, Eq.(\ref{para}) becomes the Darboux equation
\BE
\label{para2}
        ({{ \partial^2}\over{\partial \eta^2}} -              
        {{ \partial^2}\over{\partial r^2}} -              
       {{1}\over{r}} {{ \partial}\over{\partial r}})\psi=0
\EE
whose general solution \cite{courant} can be given as 
\BE
\label{darbour}
 \psi(\eta,r)= \int^1_{-1} {{f(\eta + r \mu) d\mu }\over{ \sqrt {1- \mu^2} }}
\EE
in terms of an arbitrary 
 $f=f(q)$ that is a twice-differentiable function of its
argument. Since the superluminal $\psi(\eta,r)$ are 
not standard plane-wave solutions
$\psi^{(o)}=\psi^{(o)}(z-ct)$, 
longitudinal gravitons might propagate differently from transverse
gravitons, although being both solutions of a D'Alembert wave equation.

~~~vi) on the other hand, by considering General Relativity with a cosmological
constant, it is even less clear why
 gravitons should be ordinary massless particles 
(i.e. similar to physical spin-1 photons). 
In fact, a non-zero cosmological constant $\lambda$ acts as
a graviton mass term in a linearized approximation where one replaces 
$g_{\mu \nu}= \bar{g}_{\mu \nu} + h_{\mu \nu}$, $\bar{g}_{\mu \nu}$ being a 
suitable background solution of Einstein equations with a cosmological term
$\lambda$. This observation, that to our knowledge has only been raised in 
ref.\cite{curtis}, amounts to the following.
Let us first consider the original static Einstein universe.
Using the same 
notations of \cite{pauli} (with $G_N=c^2=1$) the field equations are 
\BE
R_{ ik}-{{1}\over{2}} g_{ik}R -\lambda g_{ik} = -8\pi T_{ik}
\EE
For $T_{ik}=\mu_o\delta^i_4\delta^k_4$, where $\mu_o$ is the constant 
value of the mass density and $\lambda={{1}\over{a^2}}=4\pi \mu_o$, the
solutions are 
\BE
   \bar{g}_{ik}= \delta^i_k +{{x^ix^k}\over{ a^2 - [(x_1)^2+ (x_2)^2+(x_3)^2]}}
 ~~~~~~~(i,k=1,2,3)  
\EE
and $\bar{g}_{i4}=0$, $\bar{g}_{44}=-1$. 
 Now let us consider, in this universe, 
a point-like mass
perturbation by replacing $\mu_o \to \mu_o + M \dl$ and look for
the leading order correction $g_{44}= \bar{g}_{44}+ h_{44}$ that leads, 
 around flat space, to
$h_{44}= {{2M}\over{r}}$. Here, however, the equation for
$h_{44}$ is
\BE
          (\Delta + {{6}\over{a^2}}) h_{44} = -8\pi M \dl
\EE
Its solution is {\it not} equivalent to replace the
Newton potential with a 
Yukawa potential $\sim {{ e^{- \sqrt{\lambda} r }} \over{r}}$ 
(as Pauli says). Instead, we find an oscillatory potential 
$ { { e^{\pm ip r} }\over{r}}$, where $p^2={{6}\over{a^2}}$, and
an imaginary graviton mass $m_g=i p$ whose physical interpretation is 
unclear, as for a `wrong-sign' Klein-Gordon equation. 
Similar features are found in the G\"odel universe \cite{godel,deser} where
the presence of the cosmological term gives rise to closed time-like loops, 
or in the presently expanding-universe scenarios where 
a cosmological term is also needed to match the experimental 
observations. In the latter case, 
the linearized problem should be cast in the form
\BE
              (\hat{M}^{\alpha \beta}_{\mu \nu} ( \bar{g}) + 
\delta^\alpha_\mu \delta^\beta_\nu \lambda) 
h_{\alpha \beta}=0 
\EE
where the graviton `mass matrix' depends on the actual 
background metric and 
background energy-momentum tensor used to fit the cosmological data.
 For this reason, independently of any experiment to detect gravitational 
waves, the simple idea of massless
gravitons propagating, just as real photons, at the speed of light is far from  
being obvious. Finally, 
it was pointed out by Segal \cite{segal} that
the Einstein universe provides a very convenient covering space of Minkowski 
space-time, preserving the global causality structure of
conformal-invariant wave equations 
(Maxwell equations, massless $\lambda\Phi^4$, Yang-Mills,...). 
Therefore, at least in the case of the Einstein universe, 
the apparent unphysical nature of its gravitons may just 
indicate the failure of the linearized approximation.

~~~vii) as for the relation between the longitudinal density
fluctuations and the Newtonian potential, identifying in the
Higgs condensate genuine transverse degrees of freedom could help to
place the analogy with General Relativity on a much tighter base. 
To this end, comparing  with
superfluid  $^4$He, it would be natural to consider quantized vortices that 
can support circularly polarized transverse vibrations \cite{wilks}. 
In particular, Feynman's treatment \cite{feylow} explains very well how
vortex rings of suitable size are formed and 
can propagate almost freely, in the superfluid. This is due to 
the invariance of the superfluid wave function for a permutation 
of the atoms and does  not correspond to a physical, real circulation. 
In this sense, the remaining part of the fluid plays no role, as it would be 
for an ordinary vacuum state. This is suprisingly close to 
Lorentz's picture \cite{lorentz} of extended elementary particles that are
 "...some local modifications in the state of the aether. These
modifications may of course very well travel onward while the volume-elements
of the medium in which they exist remain at rest."
For this reason, quantized vortex rings in the Higgs condensate 
might represent the `superfluid equivalent' of 
circularly polarized transverse gravitons, just as the vortices  
invented by Thomson \cite{lord} to represent
circularly polarized transverse photons.
About the possible radius of the rings $R$ and 
their transverse dimension $d$, using Eq.(\ref{dilu}), 
there is a very wide range of $d$ and $R$, say
\BE
\label{micro}
            a \leq  d \ll R < R_{\rm mfp}
\EE
for which phions undergo no observable scattering process 
inside the ring. Therefore, following Bloch's
treatment \cite{bloch} of superfluid rings, where 
the elementary phions replace the $^4$He atoms, and considering 
the ring quantized angular momentum, 
$L=\nu \hbar$, one could try to associate circularly polarized 
graviton states to the values $\nu=\pm 2$.

~~~viii) additional connections 
with General Relativity arise when relating vortex rings to strings. To this 
end, there are two possibilities. On one hand, a tower
of vortex rings, piled along the $z$-axis and with an infinitesimal
radius in the $(x,y)$ plane, provides a 
physical representation of an open string
with a well defined angular momentum $J$ per unit length along the $z$-axis. 
When solving Einstein 
equations for such a field configuration \cite{deser}, even for zero cosmological
constant, the resulting geometry supports  
time-like loops (suitable circular paths around the string 
in the $(x,y)$ plane) thus re-proposing the problem of causality in the
presence of a scalar condensate.
On the other hand, 
a vortex ring of infinitesimal transverse section 
can also be considered a closed string whose possible excitation states, as
derived from Bloch's picture \cite{bloch}, require an infinite set of 
quantum numbers. 
This is also independent of the nature of the elementary quanta.
For instance, replacing the $^4$He atoms with Cooper pairs 
leads to a similar picture \cite{bloch}, in agreement with
the analogy between scalar and fermion superfluid vacua mentioned in the
Introduction. By considering 
superfluid rings as elementary objects, one could
try to construct an effective lagrangian and 
possible forms of ring-ring interactions. In this context, it is
interesting that the spontaneously broken phase of a
one-component $\lambda\Phi^4$ theory, in four space-time dimensions, has a 
non-trivial duality mapping \cite{leone}
into a theory of interacting membranes whose continuum limit is 
the Kalb-Ramond model \cite{kalb}. This duality transformation 
could help to connect the operators that excite the
 (st)ring-like degrees of freedom to the basic annihilation and creation 
operators for the elementary scalar quanta, in analogy with the more
conventional Bogolubov transformation from particle to phonon 
quasi-particle states.
We end up, by mentioning
that Bloch considers the possibility to account for the 
required interaction between $^4$He atoms through their mere replacement 
in the ring
by the phonon excitations of the superfluid, at least for low values 
of the associated 3-momentum where these behave as non-interacting 
quasi-particles. This means that, 
in addition to the microscopic rings of Eq.(\ref{micro}), there might be a
whole second generation of `cosmic' (st)rings, whose radial size could
extend up to the length scale associated with a free-phonon 
propagation. 

~~~ix) this last remark suggests to look for deviations from the 
free-phonon propagation that represents, 
in our picture, the mechanism for Newtonian gravity. To this end, one should
estimate the value of a mean free path associated with phonon propagation.
In superfluid $^4$He, for temperature $T\to 0$, 
the phonon mean free path becomes larger than the size of the container. 
However, in a real infinite system, one should consider its possible
effects. Notice that the phonon mean free-path is much larger than 
the phion mean free path 
$R_{\rm mfp}\sim {{1}\over{na^2}}$ that we have considered 
so far. In fact, the former depends on the {\it residual}
interactions in our infinitely diluted hard-sphere system
 where, at zero-temperature,
the phonon spectrum is almost exact. In fact, the
 residual interactions are due to a non-zero depletion $n_D$, 
the residual fraction of particles that, at zero-temperature and in the
absence of any external field, are not in the same condensed
quantum state ${\bf{p}}=0$. In the phion condensate 
where $\sqrt {na^3} = {\cal O}(10^{-16})$, this fraction 
${{n_D}\over{n}}\sim \sqrt{na^3}$  \cite{huang} is infinitesimal
and can be taken as a measure of the
residual interactions. In this sense, the relevant density 
to determine the phonon mean free path is $n_D$ and not $n$. 
Therefore, we would tentatively estimate \cite{consoli}
a phonon mean free path of 
\BE
\zeta_{\rm mfp} \sim {{1}\over{n_D a^2}} \sim 4\cdot10^{16} R_{\rm mfp}
\EE
to mark the distance over which non-linear effects 
might modify the 
free-phonon propagation and, thus, Newtonian gravity. 
For our standard value
$R_{\rm mfp}= 8$ millimeters, we find
$\zeta_{\rm mfp} \sim 3\cdot 10^{16}$cm or $2\cdot 10^3$ AU.  Checking this
prediction would 
require to detect some anomaly in the behaviour of long-period 
comets, those with $T> 200$ yr and semimajor axis larger than $\sim10^2$ AU. 
These are believed to originate from the Oort cloud \cite{oort}, a roughly
spherical condensate of  $\sim 10^{13}$ comets of average mass $\sim10^{16}$ 
grams \cite{heisler} placed 
between $10^{3.5}$ and $10^{4.5}$ AU \cite{tremaine}. 
To describe some of their
features one has to introduce some `ad hoc' assumptions \cite{tremaine}, and
this might be indicative of deviations from a pure Newtonian behaviour.
As a possible model for the deviations, we would tentatively 
follow the modification of Newtonian dynamics
(`MOND') proposed by Milgrom \cite{milgrom} as an alternative approach to the
mass discrepancy in galactic systems.  In fact, 
this is not a pure long-distance modification of gravity. Rather, it
is a modification of inertia and/or gravity that shows up in the unusual 
deep-space regime of very low acceleration, of the order of 
$g_o\sim 10^{-8}$ ${\rm cm}\cdot {\rm sec}^{-2}$. 
In this sense, MOND is the natural 
type of effect one might expect in our picture where gravity originates
as a collective oscillation of the same medium that generates inertia.
For instance, in Milgrom's approach, a long-term comet seen around the 
sun is bound \cite{milgort} when its total energy
$E={{v^2}\over{2}} -{{G_N M_{\rm sun}}\over{r}}$ is below
a condensation shell $E< \sqrt{G_N M_{\rm sun} g_o}$ 
and not only when $E <0$. 
If we attempt the identification
\BE
              g_o\sim {{G_N M_{\rm sun}}\over{ \zeta^2_{\rm mfp} }}
\EE
we obtain 
\BE             
\zeta_{\rm mfp}\sim 10^{17}~{\rm cm}
\EE
which is not too far from our reference
value $3\cdot 10^{16}$ cm. For a closer contact with Milgrom's
approach, we note that, on the base of the general arguments
 presented at the beginning of subsect.3.3, 
the deviations from free-phonon propagation 
correspond to observable changes in the phion density $n$.
These can formally be included following the treatment of quantum
liquids \cite{pita3}. In this case, the free-phonon approximation is 
equivalent to a free Lagrangian density
\BE
\label{zero}
{\cal L}_o= 
-{{1}\over{2}} (\nabla \sigma)^2 - \sigma \rho 
\EE
that upon minimization
provides the Poisson equation. Following \cite{pita3}, the residual interactions
amount to replace Eq.(\ref{zero}) with
\BE
{\cal L}= 
-{{1}\over{2}} {{n}\over{n_o}} (\nabla \sigma)^2 - {\cal W}(n) -
\sigma\rho 
\EE
i.e. allowing for $n\neq n_o$ and introducing 
an internal energy ${\cal W}(n)$. In this
case, upon minimization one finds
\BE
\label{enne}
\nabla ( {{n}\over{n_o}} \nabla \sigma)= \rho
\EE
and
\BE
\label{enne2}
     {{1}\over{2n_o}} (\nabla\sigma)^2 + {{d {\cal W} }\over{d n}}=0 
\EE
Finally, after deriving $n=n( |\nabla\sigma|)$ from
Eq.(\ref{enne2}), one has to introduce a constant acceleration $g_o$ 
in order the function $\mu= {{n}\over{n_o}}$ to be dimensionless. 
Therefore, using typical techniques of quantum field theories \cite{piran},
one may attempt a microscopic derivation of Milgrom's function 
$\mu=\mu({{|\nabla \phi|}\over{g_o}})$ 
used in his non-linear description of gravity
\BE
\nabla(\mu({{|\nabla \phi|}\over{g_o}})\nabla \phi)=\rho
\EE
In fact, as discussed
in ref.\cite{beke}, this has the same form of the stationary flow of
an irrotational fluid of suitable density. However, it can also be interpreted
as the gravitational analog of a dielectric medium where 
$\nabla (\epsilon (|{\bf{E}}|) {\bf{E}})= \rho_e$ with 
${\bf{E}}=-\nabla U$, $U$ being the electrical potential. This type of 
correspondence is very natural in our approach where gravity originates from
the density fluctuations of the scalar condensate.

\vskip 20 pt
{\bf Acknowledgements}~~~I thank J. H. Field, A. Garuccio, E. Giannetto, 
A. Pagano, E. Recami, M. Roncadelli, G. Salesi, F. Selleri, P. M. Stevenson
and D. Zappal\'a for useful discussions.

\end{document}